\documentclass[prb, twocolumn,  amsmath, amssymb, showpacs,  floatfix]{revtex4}
\usepackage{amsmath,graphics, graphicx, bm, natbib, hyperref}
\usepackage[usenames,dvipsnames]{color}
\usepackage[lofdepth,lotdepth, caption=false]{subfig}

\begin{document}

\author{Wijnand Broer, George Palasantzas, and Jasper Knoester}
\affiliation{Zernike Institute for Advanced Materials, University of Groningen, \\
Nijenborgh 4, 9747 AG Groningen, the Netherlands}

\author{Vitaly B. Svetovoy} 
\affiliation{MESA\textsuperscript{+} Institute for Nanotechnology, University of Twente, \\
P.O. Box 217, 7500 AE Enschede, the Netherlands}
\affiliation{{Institute of Physics and Technology, Yaroslavl Branch, Russian Academy of Sciences, 150007 Yaroslavl, Russia.}}

\title {Significance of the Casimir force and surface
roughness for actuation dynamics of MEMS}

\date{\today}

\begin{abstract}
Using the measured optical response and surface roughness topography as inputs, we perform realistic calculations of the combined effect of Casimir and electrostatic forces on the actuation dynamics of micro-electromechanical systems (MEMS). In contrast with the expectations, roughness can influence MEMS dynamics even at distances between bodies significantly larger than the root-mean-square roughness. This effect is associated with statistically rare high asperities that can be locally close to the point of contact. It is found that, even though surface roughness appears to have a detrimental effect on the availability of stable equilibria, it ensures that those equilibria can be reached more easily than in the case of flat surfaces. Hence our findings play a principal role for the stability of microdevices such as vibration sensors, switches, and other related MEM architectures operating at distances below 100 nm.\end{abstract}

\maketitle

\section{Introduction}
Electromagnetic fluctuations that pervade any medium including empty space generate forces between neutral bodies known as Casimir-Lifshitz forces, of which van der Waals forces are special cases.\cite{Casimir48, Lifshitz55,Lifshitz61, Israelachvili, London37, Ninhambook, vdWBook} Casimir forces arise from electromagnetic waves created by quantum and thermal fluctuations. \cite{Casimir48,Lifshitz55,Lifshitz61, Israelachvili, London37, Ninhambook, vdWBook, Johnson2011, Lambrecht2002, Kardar99, Lamoreaux2005, Lamoreaux2007, MilonniBook, MiltonBook, Milton2004, Onofrio2006, Plunien86, Spruch96, Rodriguez2008, Miri2008, Genet2008, ParallelPlates2002} These are expected to become important as components of micro-electro-mechanical systems (MEMS) enter submicrometer separations.\cite{Lamoreaux2005, Ball2007, CapassoReview2007, GeorgePhaseMap, GeorgePullIn, BuksEPL2001, Serry98, Serry95, DelRio2005, OsterbergThesis, BochobzaDegani2002A, BochobzaDegani2002B, Zhang2002, Lin2003, Esquivel2005, Esquivel2009, LinCSF2005, LinMST2005} The small scales at which MEM engineering is now conducted have revitalized interest in the Casimir force since devices such as vibration sensors and switches are made with parts that are just a few micrometers in size and have the right size for the Casimir force to play a role: they have surface areas big enough but gaps small enough for the force to draw components together and lock them tight, which is an effect called stiction. In fact, permanent adhesion is a common cause of malfunction in MEMS devices. Casimir forces, in synergy with electrostatic actuating forces, can further augment this phenomenon.\cite{Ball2007, CapassoReview2007, GeorgePhaseMap, GeorgePullIn, BuksEPL2001, Serry98, Serry95, DelRio2005, OsterbergThesis, BochobzaDegani2002A, BochobzaDegani2002B, Zhang2002, Lin2003, Esquivel2005, Esquivel2009, LinCSF2005, LinMST2005} Additionally, as the development of MEMS evolves toward nanomechanical systems (NEMS), attention will also be drawn to scaling issues. It is inevitable that the Casimir interactions between metallic and/or dielectric surfaces in nanometer proximity of each other will occur, and stiction phenomena require specific attention. On the other hand, the irreversible adhesion of moving parts resulting from Casimir and electrostatic forces, \cite{Ball2007, CapassoReview2007, GeorgePhaseMap, GeorgePullIn, BuksEPL2001, Serry98, Serry95, DelRio2005, OsterbergThesis, BochobzaDegani2002A, BochobzaDegani2002B, Zhang2002, Lin2003, Esquivel2005, Esquivel2009, LinCSF2005, LinMST2005} can also be exploited to add new functionalities to micromechanical architectures. 

\begin{figure*}[!hptb]		\subfloat[][]{\includegraphics[width=0.45\textwidth]{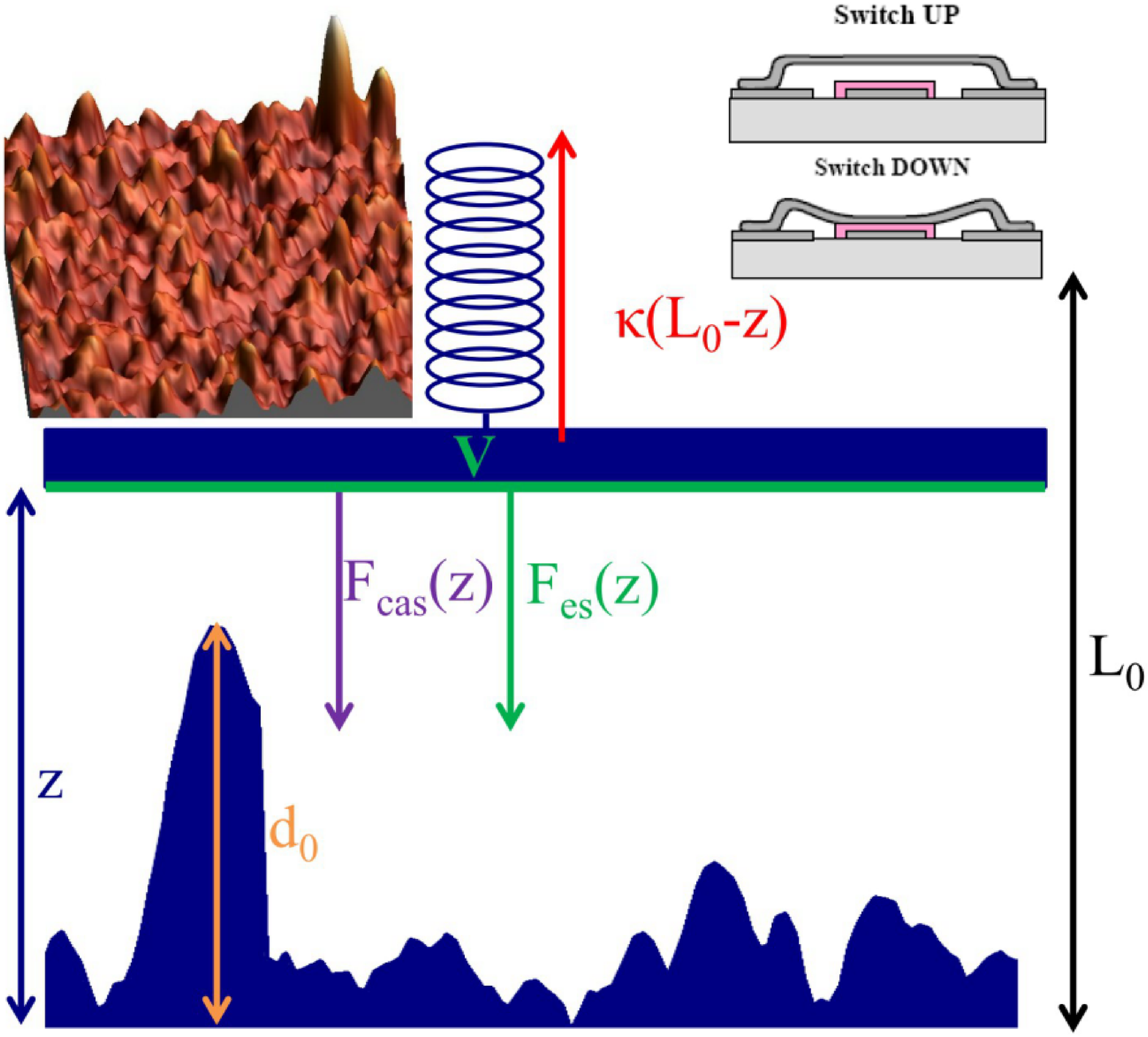}
	\label{fig:fig1a}}
\subfloat[][]{\includegraphics[width=0.45\textwidth]{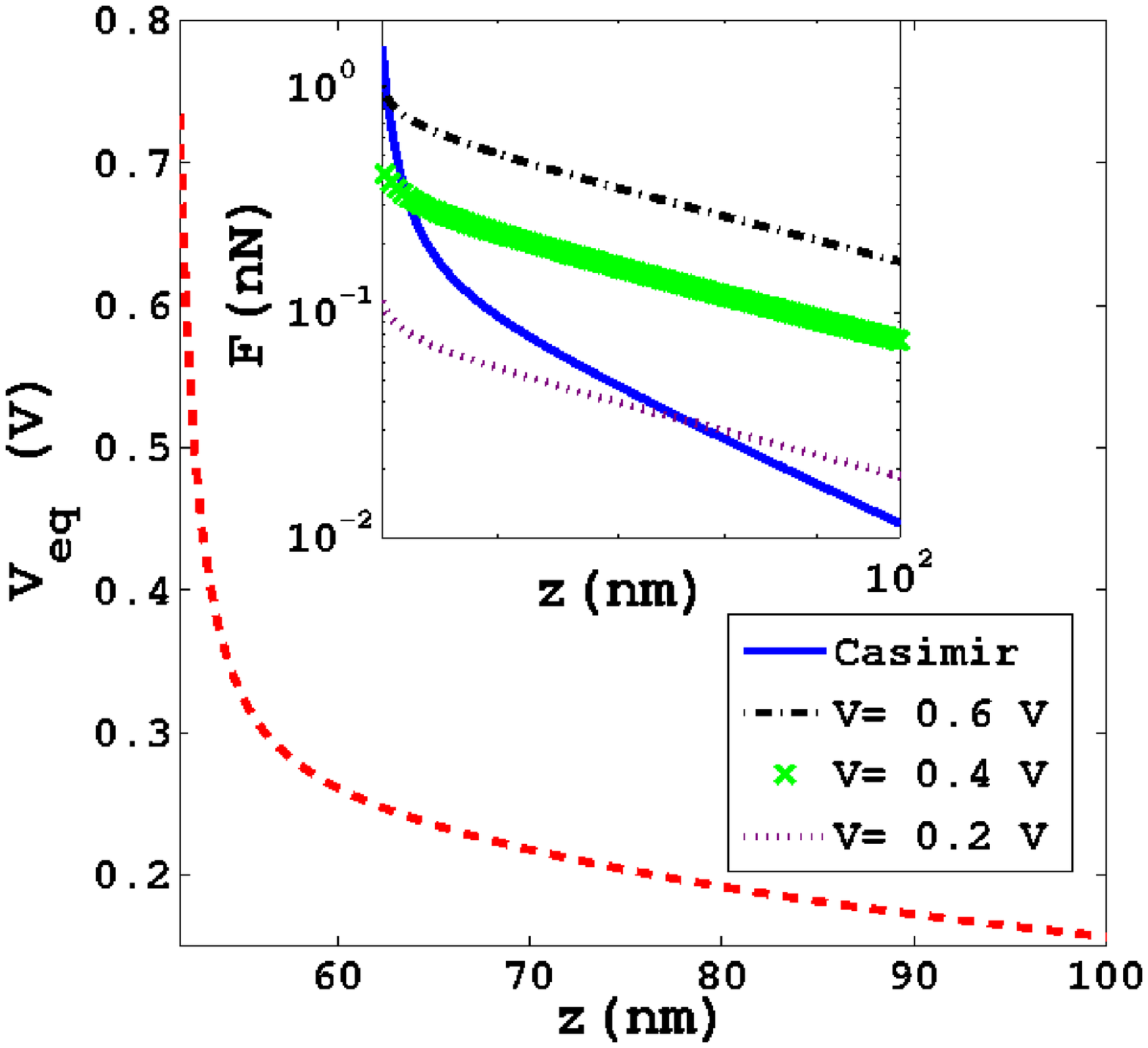}
	\label{fig:fig1b}}	
\caption{\protect\subref{fig:fig1a} Schematic of the MEM system. The spring tries to move the oscillator toward the initial separation $L_0$. The contact distance $d_0$ is the maximum height of the asperities on a rough surface. \protect\subref{fig:fig1b} In the inset, the Casimir force from Eq. \eqref{eq:Fcas} for gold surfaces is compared to the electrostatic force between rough surfaces from Eq. \eqref{eq:Fetot} for several values of the applied voltage $V$. The quantity $V_{eq}$ indicates to what voltage the Casimir corresponds after equating the Casimir force from Eq. \eqref{eq:Fcas} to the electrostatic force from Eq. \eqref{eq:Fetot}. In this case the r.m.s. $w=10$ nm  and the contact distance $d_0=$ 50.8 nm.}
\label{fig:fig1}
\end{figure*}

Casimir forces that are responsible for stiction in dry conditions,\cite{DelRio2005} and thus profoundly influence the actuation dynamics, supplement the electrostatic force in countering the elastic restoring force to determine, for example, the beam's actuation behavior in microswitches.  The latter is typically constructed from two conducting electrodes of which one is fixed and the other is suspended by a mechanical spring governed by Hooke's law\cite{PeleskoBernstein} (see Fig. \ref{fig:fig1a}). Voltage application between the electrodes moves the electrodes toward each other because of the electrostatic force. At a certain voltage, the moving electrode becomes unstable and collapses or pulls-in onto the ground electrode.\cite{OsterbergThesis, BochobzaDegani2002A} Residual stress and fringing field effects have also shown to have a great influence on the behavior of RF (radio frequency) switches, and strongly influence their failure characteristics.\cite{BochobzaDegani2002B, Zhang2002} 

In earlier investigations of the effect of the Casimir or van der Waals forces on the dynamical behavior of nanoscale electrostatic actuators, roughness was either ignored or only weak roughness was considered. In some cases tabulated optical data were taken into account.\cite{BuksEPL2001, Serry98, Serry95} In this paper, we will explore the actuation dynamics of microswitches made from real materials (with a definite measured optical response \cite{PeterOptical, AISTPRA, AISTADVMAT} and characterized by some degree of nanoscale roughness) accounting for both electrostatic and Casimir forces, which counteract an elastic restoring force (see Fig. \ref{fig:fig1a}). Advances made in the measurement and theoretical understanding of Casmir forces over the last 10 years allow today a more detailed study of MEMS made from real material surfaces.\cite{BroerEpl2011, BroerPRB2012, Zhao99} Note that although electrostatic forces can be switched off if no potential is applied, Casimir forces will always be present and will influence the actuation dynamics. 

\section{Roughness correction to the electrostatic force}
It has been shown\cite{BroerEpl2011, BroerPRB2012} that the disagreement between the experiment\cite{PeterRoughness} and the theory describing roughness perturbatively\cite{Genet2003, Lambrecht2005} can be resolved by taking into account rare high peaks on rough surfaces. These high peaks can be described by ``extreme value statistics'' as follows from a statistical analysis of AFM topography data for gold films.\cite{BroerEpl2011, BroerPRB2012, d0paper} Indeed, recently there has been more awareness of the importance of extreme value statistics for the analysis of rough surfaces.\cite{Lu2012}
The Casimir force between rough surfaces can be written as:
\begin{equation}\label{eq:Fcas}
F_{Cas}(z)=F_{PT}(z)+F_{peaks}(z).
\end{equation}
The term $F_{PT}(z)$ denotes the Casimir force between rough surfaces from Ref. \onlinecite{Genet2003}, which includes a ``zeroth'' order contribution  corresponding to flat surfaces and a perturbative roughness correction $\sim{}(w/z)^2$, where $w$ is the root-mean-square roughness. The term $F_{peaks}(z)$ represents the contribution of high peaks, which is associated with the aforementioned extreme value statistics. It is important to note that $F_{Cas}(z)$ is singular at the distance upon contact,\cite{d0paper, BroerEpl2011, BroerPRB2012} ($z=d_0$) which is the real minimum separation due to surface roughness. It is assumed that the contributions of high peaks are independent of each other, an approximation justified by the large horizontal distance between them. This distance is large because such peaks are statistically rare events.\cite{BroerEpl2011, BroerPRB2012} The ellipsometry measurements for gold samples reported in Ref. \onlinecite{PeterOptical} were used as optical data.

The roughness correction to the electrostatic force can be obtained in the same way as was done in Ref. \onlinecite{BroerEpl2011} for the Casimir force: the heights of the surface comparable to $w$ can be taken into account perturbatively, whereas the contribution of high peaks can be approximated by treating each peak independently. This approximation is justified by the large distance between the high peaks, because  such peaks are statistically rare events. The perturbative roughness correction to the electrostatic force is based on an analysis for isotropic roughness,\cite{Zhao99,  GeorgePhaseMap, GeorgePullIn} as is the case for the gold films considered here\cite{PeterOptical} which are grown under non equilibrium conditions. This correction starts by modeling the surfaces as a capacitor with capacitance:
\begin{eqnarray}\label{eq:capacitance}
& \left\langle C(z) \right\rangle  =   \nonumber\\      &  \frac{A\varepsilon}{z}\Big\{1+\frac{2(2\pi)^4}{A}\int\limits_{0}^{k_c} \left\langle|\tilde{h}(\mathbf{k})|^2\right\rangle\big(k^2
+\frac{\pi}{z}\coth(kz)\big)\mathrm{d}\mathbf{k}\Big\} 
\end{eqnarray}
where the first and second terms correspond to flat surfaces and a second order perturbative correction, respectively, and $A$ denotes the surface area of each plate. The quantity $k_c$ represents the wavenumber corresponding to a lower lateral roughness cutoff of the order of the inter-atomic distances ($\sim4$ \AA{} for gold) . For power law or self affine random roughness a suitable model for the power spectrum  $\left\langle|\tilde{h}(\mathbf{k})|^2\right\rangle$  to perform calculations with is given by\cite{George93}
\begin{equation}\label{eq:spectrum}
\left\langle|\tilde{h}(\mathbf{k})|^2\right\rangle = \frac{A}{(2\pi)^5}\frac{w^2\xi^2}{(1+ak^2\xi^2)^{1+H}}.
\end{equation}
Here $\xi$ denotes the correlation length, $a$ represents the self-affine roughness parameter which can be found by solving the algebraic equation: $a=1/(2H)[1-(1+ak_c^2\xi^2)^{-H}]$. For the gold films considered here\cite{PeterOptical} the roughness exponent has the value $H=0.9$.

The electrostatic force including the second order perturbative roughness correction can be written as
\begin{equation}\label{eq:Fpe}
F_{pe}(z)=-\tfrac{1}{2}V^2\frac{\mathrm{d}\left\langle C(z) \right\rangle}{\mathrm{d}z},
\end{equation}
where the average capacitance $\left\langle C(z) \right\rangle$ is given by Eq. \eqref{eq:capacitance} and $V$ denotes the applied voltage. Equipotential planes are expected to be a valid approximation at separations below 100 nm. Statistical deviations from this approximation, known as potential patches, typically play a role at separations of the order of a few hundred nanometers up to a few micrometers.\cite{Sushkov2011, Behunin2012a, Behunin2012b} Now the contribution of the high peaks in the surface can be approximated by a sum of separate contributions of each peak as it was done in Ref. \onlinecite{BroerEpl2011} for the Casimir force. For this purpose we start with the electrostatic force between flat surfaces:
\begin{equation}\label{eq:Fe}
F_e(z)\cong\frac{\varepsilon_0 AV^2}{2z^2}.
\end{equation} 
For the roughness statistics we can use the same AFM topography data with the same statistical analysis as in Ref. \onlinecite{BroerEpl2011}. Therefore the electrostatic force between rough surfaces becomes
\begin{eqnarray}\label{eq:Fetot}
& F_{es}(z) = F_{pe}(z) + \\\nonumber  & \int\limits_{d_1}^{d_0}{f(d)[F_e(z-d)-F_e(z)+dF_{e}'(z)-\tfrac{1}{2}d^2F_e''(z)]\mathrm{d}d} & \\\nonumber
\end{eqnarray}
where $f(d)$ denotes the probability density function. The height $d_1=3w$ is the separation above which $f(d)$ can be fitted to a Gumbel distribution\cite{BroerEpl2011, BroerPRB2012} and $d_0$ is the height of the highest asperity (see Fig. \ref{fig:fig1a}). It must be noted that the expression in Eq. \eqref{eq:Fetot} is also singular at $z=d_0$.

Figure \ref{fig:fig1b} shows the relative strength of electrostatic and Casimir forces for various potentials between real nanoscale rough Au-Au surfaces. The Casimir force becomes significant for separations $z<100$ nm and overcomes the electrostatic force rather rapidly as the applied potential drops below 1 V (a regime typical for MEMS) and separations close to distance upon contact due to surface roughness. Indeed, the potential $V_{eq}(z)$ where $F_{Cas}(z)=F_{es}(z)$ increases rapidly toward smaller separations, which shows that the Casimir force corresponds to increasing values of the applied voltage $V$. These results clearly show that below 100 nm Casimir forces can strongly influence the actuation dynamics.

\section{Actuation dynamics of MEMS}
Modeling a MEMS as a classical mass-spring system has been well established.\cite{PeleskoBernstein}  Let the separation $z$ depend on time and satisfy the following differential equation:
\begin{equation}\label{eq:EOM}
m\frac{\mathrm{d}^2z}{\mathrm{d}t^2}=\kappa(L_0 - z)-F(z) ,
\end{equation}
where $F(z)=F_{Cas}(z)+F_{es}(z)$ represents the total surface force, $\kappa$ is the spring constant, and $L_0$ is the distance between bodies if no external force is present, $F(z)=0$. The effective mass $m$ merely rescales the equation \eqref{eq:EOM}. For our calculations we used as an example  a resonance frequency $\omega_0\equiv\sqrt{\kappa/m}=300\cdot2\pi$ rad/s which is typical for a wide variety of commercial resonators, e.g. tapping mode AFM cantilevers and other doubly clamped beam MEMS\cite{Garcia2002}. This frequency is kept constant, whereas $\kappa$ is varied and used as a control parameter.

The solutions of Eq. \eqref{eq:EOM} can be investigated with a \emph{phase portrait},\cite{HirschSmaleDevaney} i.e. a plot of $z$ versus $z'(t)$. Studies of the influence of the Casimir force for nanoscale electrostatic actuators with flat, perfectly conducting electrodes showed that their phase portraits exhibit periodic orbits around a center equilibrium, and an orbit that passes through an unstable saddle point.\cite{LinCSF2005, LinMST2005}  These studies were extended to the influence of weak roughness only. \cite{GeorgePhaseMap, GeorgePullIn}

Consider first the case of zero electrostatic force, $F_{es}(z)=0$. The goal is to find out under what conditions the oscillator described by Eq. \eqref{eq:EOM} can return to its original position; i.e. for what parameter values periodic solutions exist. \emph{The existence of periodic solutions indicates that the spring is strong enough to prevent stiction.} If the spring constant is large enough, the stable center around which periodic solutions exist will be accompanied by an unstable saddle point equilibrium.\cite{LinCSF2005, LinMST2005, Esquivel2005, Esquivel2009} If the spring constant becomes lower, the center and saddle point will merge into an unstable ``center-saddle'' point. For an even lower value of $\kappa$ there are no equilibria at all. This is an example of what is known as a saddle-node bifurcation.\cite{HirschSmaleDevaney} 

In order to understand for what values of $\kappa$ such equilibrium points are available we introduce the following bifurcation parameter
\begin{equation}\label{eq:bif1}
\lambda_{cas}\equiv\frac{F_{L}(L_0)}{\kappa L_0},
\end{equation}
where $F_{L}(L_0)$ denotes the Casimir force given by the Lifshitz formula\cite{Lifshitz55,Lifshitz61} (for flat surfaces) at $z=L_0$. This ratio of the minimal Casimir force and the maximal elastic restoring force represents the relative importance of one force compared to the other. In an equilibrium the total force  given by Eq. (\ref{eq:EOM}),  is zero:  $F_{tot}(z)=0$. This case yields $\lambda_{cas}=(1-z_s/L_0)F_L(L_0)/F_{cas}(z_s)$, where $z_s$ denotes the locus of the stationary points.

The results are plotted in Fig. \ref{fig:fig2a}. As one can see the rougher the sample (i.e. the higher the value of the contact distance\cite{d0paper} $d_0$), the higher the spring constant must be to get equilibria and periodic solutions. The maximum of $\lambda_{cas}$ decreases with $d_0$. The \emph{position} of this maximum changes only slightly under the influence of random roughness: from $0.78L_0$ for a flat surface to $0.81L_0$ for the roughest sample. This is because at these separations the roughness effect is small (perturbative) and does not drastically change the force. To clarify the meaning of Fig. \ref{fig:fig2a}, the general solution, represented by the phase portrait, is plotted for three different values of the spring constant, for the roughest sample (with $d_0=$ 50.8 nm). Fig. \ref{fig:fig2b} shows the case where the spring constant is large enough for the bifurcation parameter to be below its maximum value. In this case  there are two equilibria: the stationary point closest to $L_0$ is a (stable) center around which periodic solutions (closed curves) exist. Since the system considered here is conservative (Hamiltonian), the phase portraits can be obtained by plotting the level curves of the total energy.  The solutions of Eq. \eqref{eq:EOM} are periodic if the amplitude stays below a value of approximately 0.4$L_0$.  The shift of the minimum separation due to roughness from zero to $d_0$ prevents periodic motion if the total energy is too high. In Fig. \ref{fig:fig2c} the value of $\kappa$ has been chosen such that it corresponds to the maximum value of $\lambda_{cas}$ in Fig. \ref{fig:fig2a}. In this case there is only one equilibrium, known as a center-saddle point,\cite{Hanssmann} which are always unstable. There are no periodic solutions in this case. If the value of the spring constant is lowered further, no equilibria are available anymore. The spring is too weak to counterbalance the attractive Casimir force. The solution for this case is plotted in Fig. \ref{fig:fig2d}.

Although neglecting the electrostatic force can provide some insight, this force must also be taken into account. If we consider the presence of the electrostatic force only, $F_{es}(z)\neq0$ and $F_{Cas}(z)=0$, we can define an additional bifurcation parameter for the electrostatic force:
\begin{equation}\label{eq:bif2}
\lambda_{es}\equiv\frac{\varepsilon_0AV^2}{2\kappa L_0^3}=\frac{F_e(L_0)}{\kappa L_0}.
\end{equation}
Similarly to the previous case, this is the ratio of the minimum electrostatic force and the maximum elastic force, which is a measure of the relative importance of one force compared to the other. In this case it holds that $\lambda_{es}=F_e(L_0)/F_{es}(z_s)(1-z_s/L_0)$, which is obtained from the condition $F_{tot}=0$. However, it must be stressed that this case is a rather artificial one, because the Casimir force cannot be shut down (since it stems from quantum mechanical uncertainty). Results for this case are qualitatively similar to the previous one, but the roughness effect is less pronounced because the electrostatic force depends more weakly on the separation distance than the Casimir force.

In  the more general case $F_{es}(z)\neq0$ and $F_{Cas}(z)\neq0$, the stationary points $z_s$ satisfy the following equation obtained from the condition $F_{tot}=0$:
\begin{equation}\label{eq:3dbif}
1-\frac{z_s}{L_0}-\frac{F_{cas}(z_s)}{F_{L}(L_0)}\lambda_{cas}-\frac{F_{es}(z_s)}{F_e(L_0)}\lambda_{es}=0,
\end{equation}
 where $\lambda_{cas}$ and $\lambda_{es}$  are defined by Eqs. \eqref{eq:bif1} and \eqref{eq:bif2} respectively. Eq. \eqref{eq:3dbif} is an implicit function of two variables, plotted in Fig. \ref{fig:fig3a} for both the idealized case of flat surfaces and the roughest surface with $d_0= 50.8$ nm. The  graph for the case of flat surfaces encloses the one for the rough surface case. This indicates that, similarly to the previous case, surface roughness increases the minimum value of the spring constant required to compensate for both the electrostatic and Casimir forces. However in this case this value further increases with the value of the applied voltage $V$, as indicated by the contour plots for the rough case in Figs. \ref{fig:fig3b} and \ref{fig:fig3c}. The phase portraits are similar to those in Fig. \ref{fig:fig2}: the only difference is that the distance between the center and saddle point is smaller than in Fig. \ref{fig:fig2a}, since $V\neq 0$ in this case. From Fig. \ref{fig:fig3b} it can be concluded that this distance decreases with $V$ since $\lambda_{es}\propto{}V^2$. 
 
The critical equilibrium points at which stiction occurs are characterized by the conditions $dU/dz=0$ and $d^2U/dz^2=0$, where $U$ denotes the total potential energy of the system (i.e. $F_{tot}(z)=-dU/dz$). See e.g. Refs. \onlinecite{Esquivel2005}, \onlinecite{LinCSF2005}, or \onlinecite{PeleskoBernstein}. Hence Eq. \eqref{eq:3dbif} must be combined with its derivative with respect to $z_s$, which is also set to zero:
\begin{equation}\label{eq:3dbif2}
-\frac{1}{L_0}-\frac{F_{cas}'(z_s)}{F_{L}(L_0)}\lambda_{cas}-\frac{F_{es}'(z_s)}{F_e(L_0)}\lambda_{es}=0,
\end{equation}
which corresponds to $d^2U/dz^2=-dF_{tot}/dz=0$. Eqs. \eqref{eq:3dbif} and \eqref{eq:3dbif2} form a system of two equations with three unknowns. Eliminating $z_s$ from this system yields a relation between the bifurcation parameters $\lambda_{cas}$ and $\lambda_{es}$. This relation is plotted in Fig. \ref{fig:fig3d}. Since $\lambda_{es}\propto V^2/\kappa$ and $\lambda_{cas}\propto1/\kappa$ the origin of this graph corresponds to high spring constants and low voltages, i.e. to the situation that the spring is strong and the device is shut down (i.e. $V$ decreases such that $\lambda_{es}\ll1$). In Fig. \ref{fig:fig3d} it can be seen that the equilibria join in the origin both for the flat and the rough case. This corresponds to the ideal case where only periodic motion exists and stiction is of no concern. It turns out that this limit is reached \emph{faster} as a function of the parameter values in the rough case than in the flat one. On the other hand, stable and unstable equilbria are closer in the rough than in the flat case.  It should be kept in mind, however, that the rough case in Fig. \ref{fig:fig3d} corresponds to the roughest sample available with $d_0=50.8$ nm. Presumably the optimum in terms of preventing stiction lies somewhere between these two cases (e.g. a contact distance between 20 nm and 30 nm). 
 
\section{Conclusions}
It may seem surprising that a flat surface is not optimal in terms of preventing stiction, but the reason for this is that the shift of the minimum separation from zero to $d_0$ prevents both the Casimir and electrostatic forces from becoming too large. In other words, there is a trade-off between two effects: on one hand random surface roughness requires a higher spring constant for equilibria to exist and puts the stable and unstable equilibria closer together, and on the other hand it prevents the surfaces from reaching too short a separation, reducing both the Casimir and electrostatic forces.

\begin{figure*}[!hptb]
	\centering		
\subfloat[][]{\includegraphics[width=0.45\textwidth]{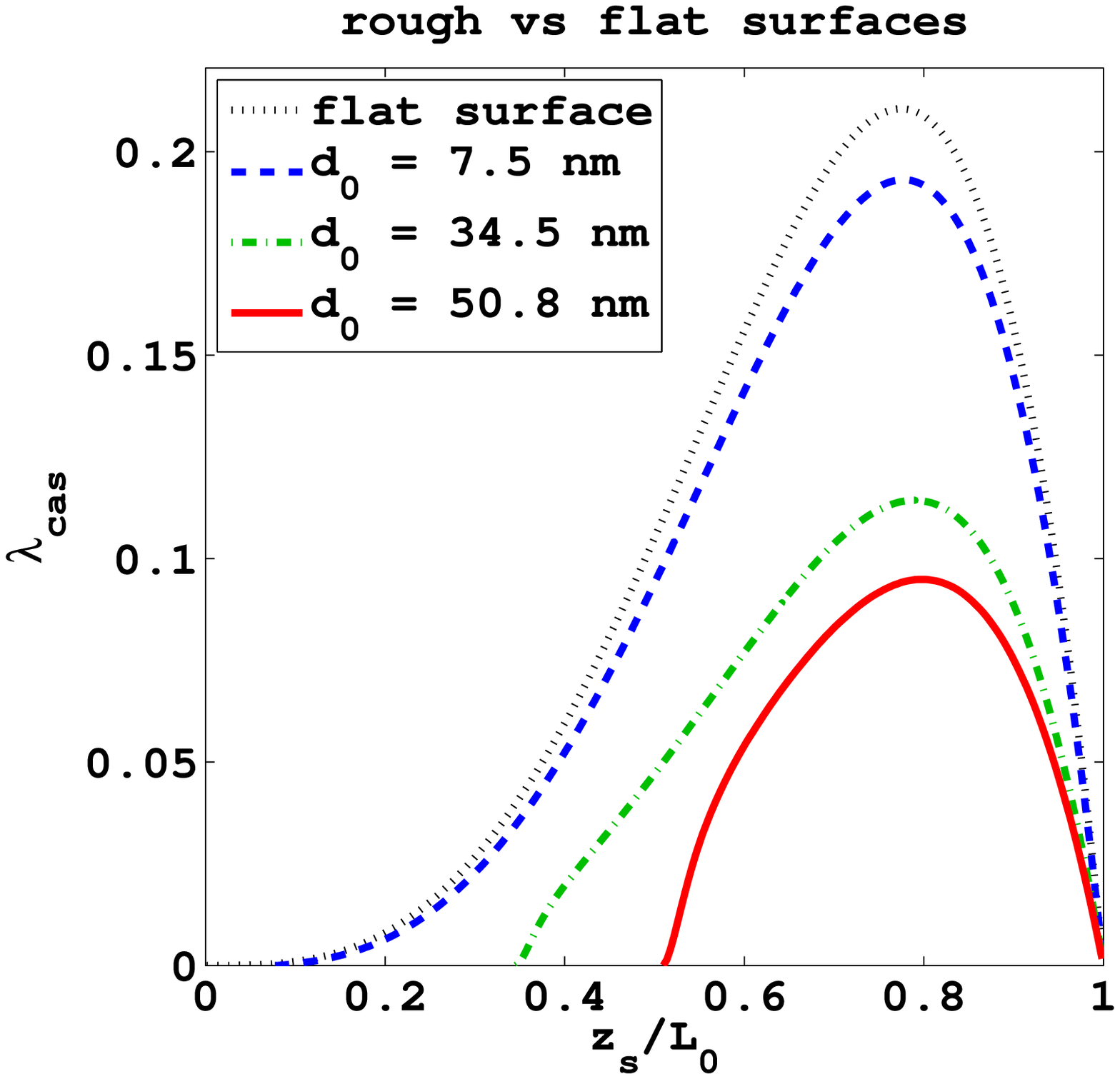}\label{fig:fig2a}}
\subfloat[][]{\includegraphics[width=0.45\textwidth]{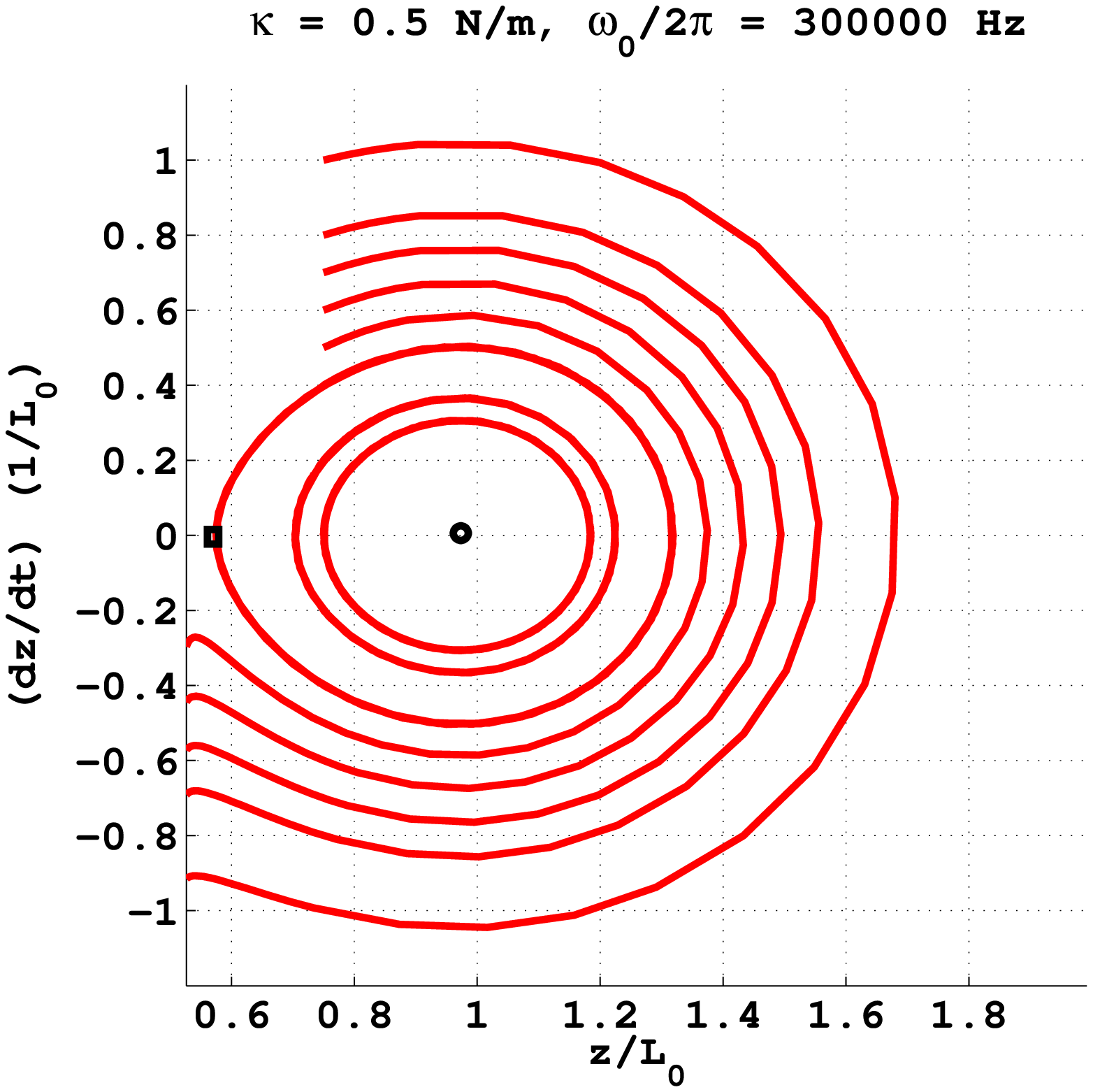}\label{fig:fig2b}}\\
\subfloat[][]{\includegraphics[width=0.45\textwidth]{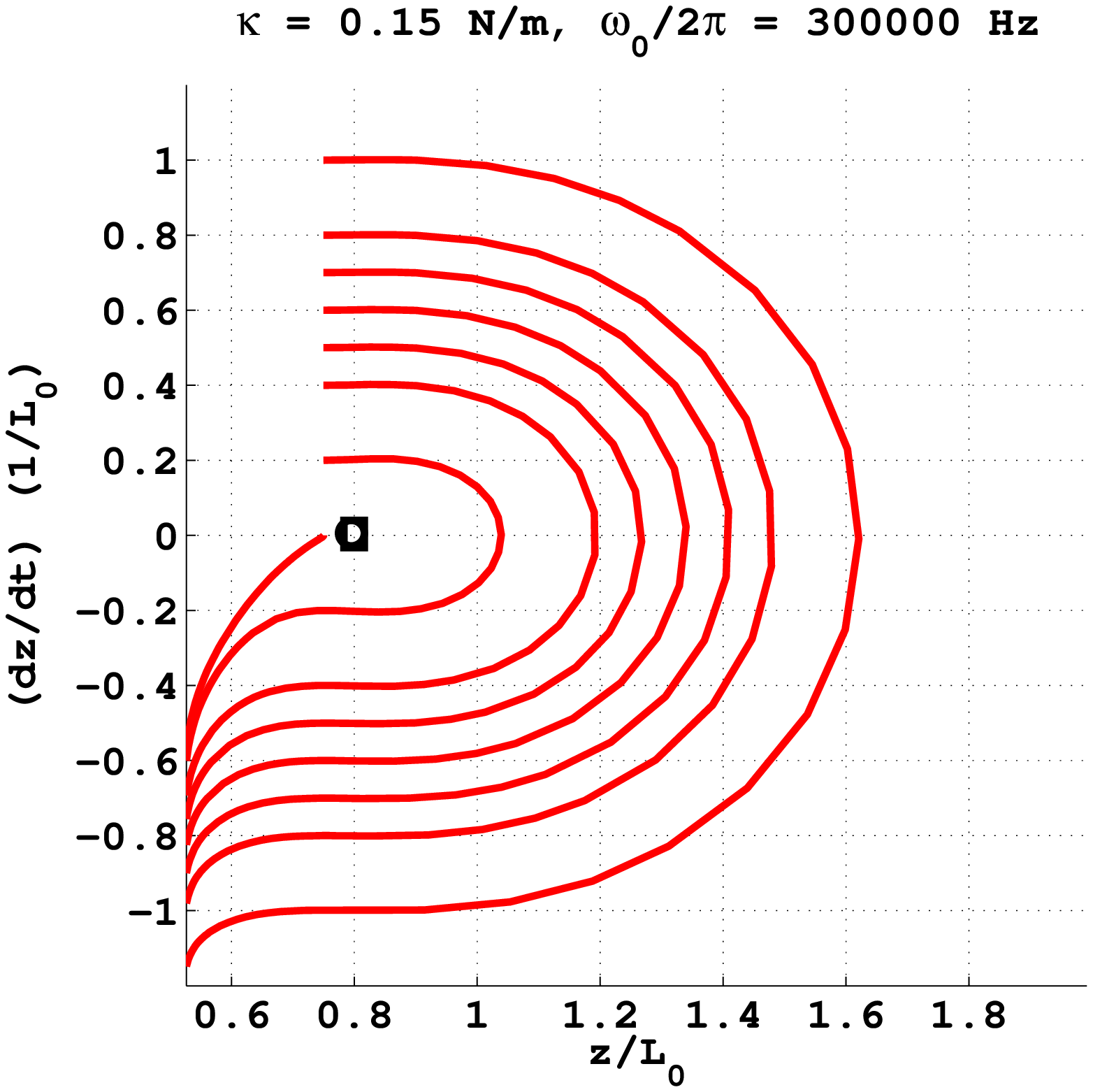}\label{fig:fig2c}}
\subfloat[][]{\includegraphics[width=0.45\textwidth]{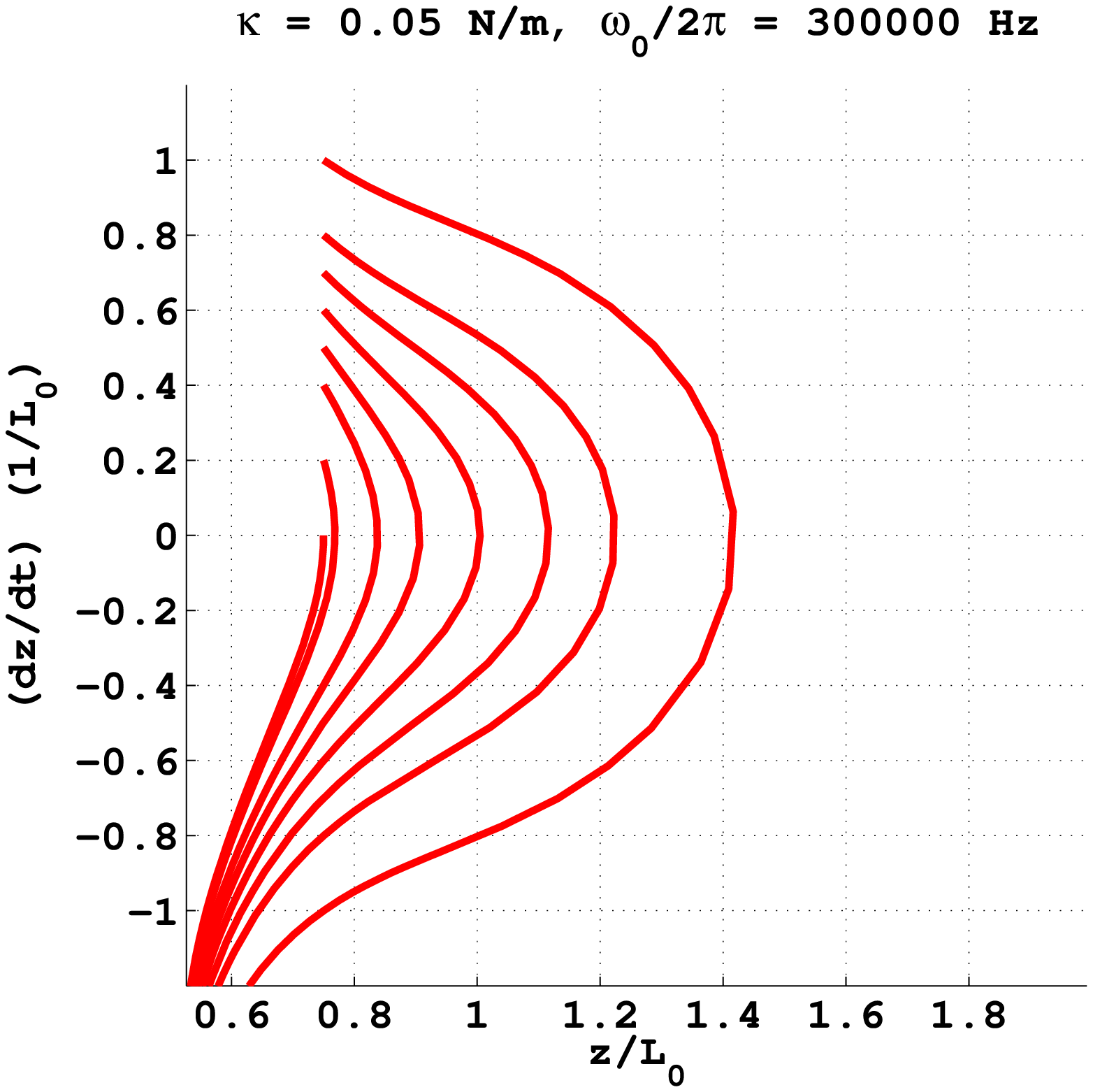}\label{fig:fig2d}}
	\caption{ \protect\subref{fig:fig2a} Bifurcation diagram of Casimir actuated MEMS for various rough surfaces, each of which is associated with a different value of the contact distance $d_0$. The value of the initial separation $L_0=$ 100 nm.  \protect\subref{fig:fig2b} Phase portrait for the surface with $d_0=$ 50.8 nm. The (black) circle and the (black) square indicate the positions of the center and saddle point equilibria, respectively. In this case, the spring constant is high enough for periodic solutions to exist. \protect\subref{fig:fig2c} This phase portrait corresponds to the maximum of the solid (red) curve in Fig. \ref{fig:fig2a}. In this case there is only one (unstable) equilibrium, and there are no periodic solutions.  \protect\subref{fig:fig2d} Phase portrait corresponding to a point above the maximum of solid (red) curve in Fig. \ref{fig:fig2a}. There are no equilibria in this case. }
\label{fig:fig2}	
\end{figure*}

\begin{figure*}[!hptb]
	\centering
		\subfloat[][]{\label{fig:fig3a}\includegraphics[width=0.45\textwidth]{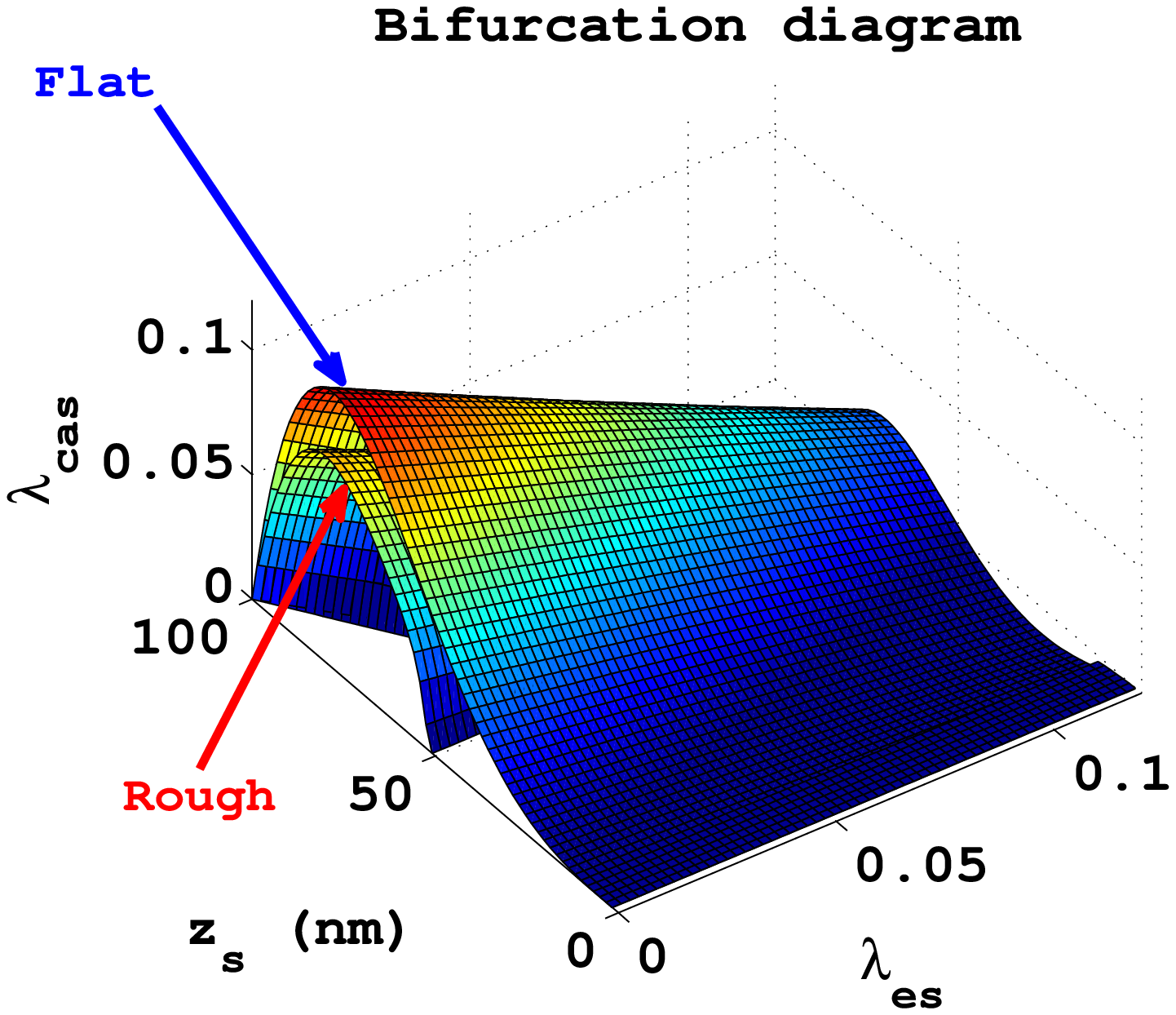}}
\subfloat[][]{\label{fig:fig3b}\includegraphics[width=0.45\textwidth]{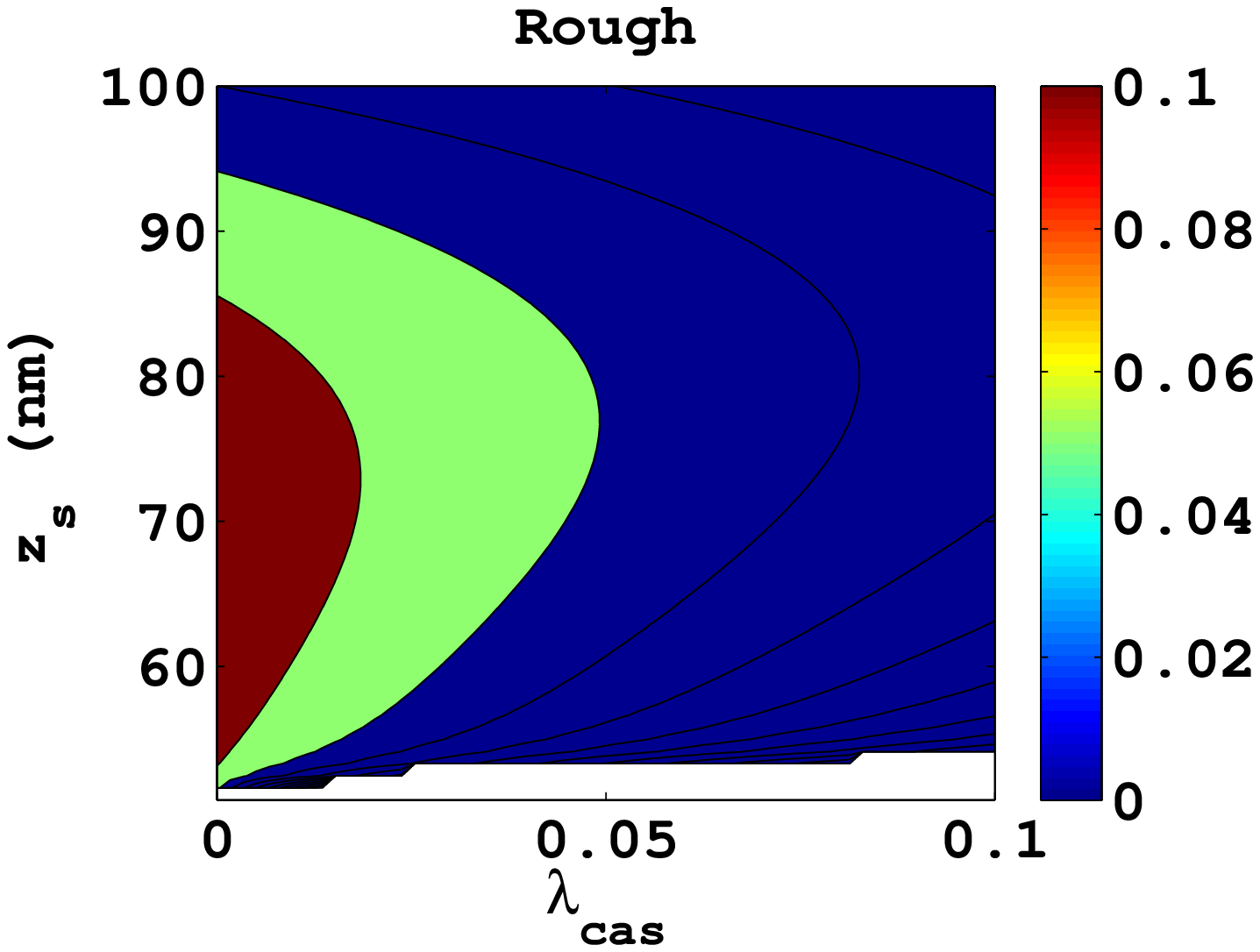}}\\
\subfloat[][]{\label{fig:fig3c}\includegraphics[width=0.45\textwidth]{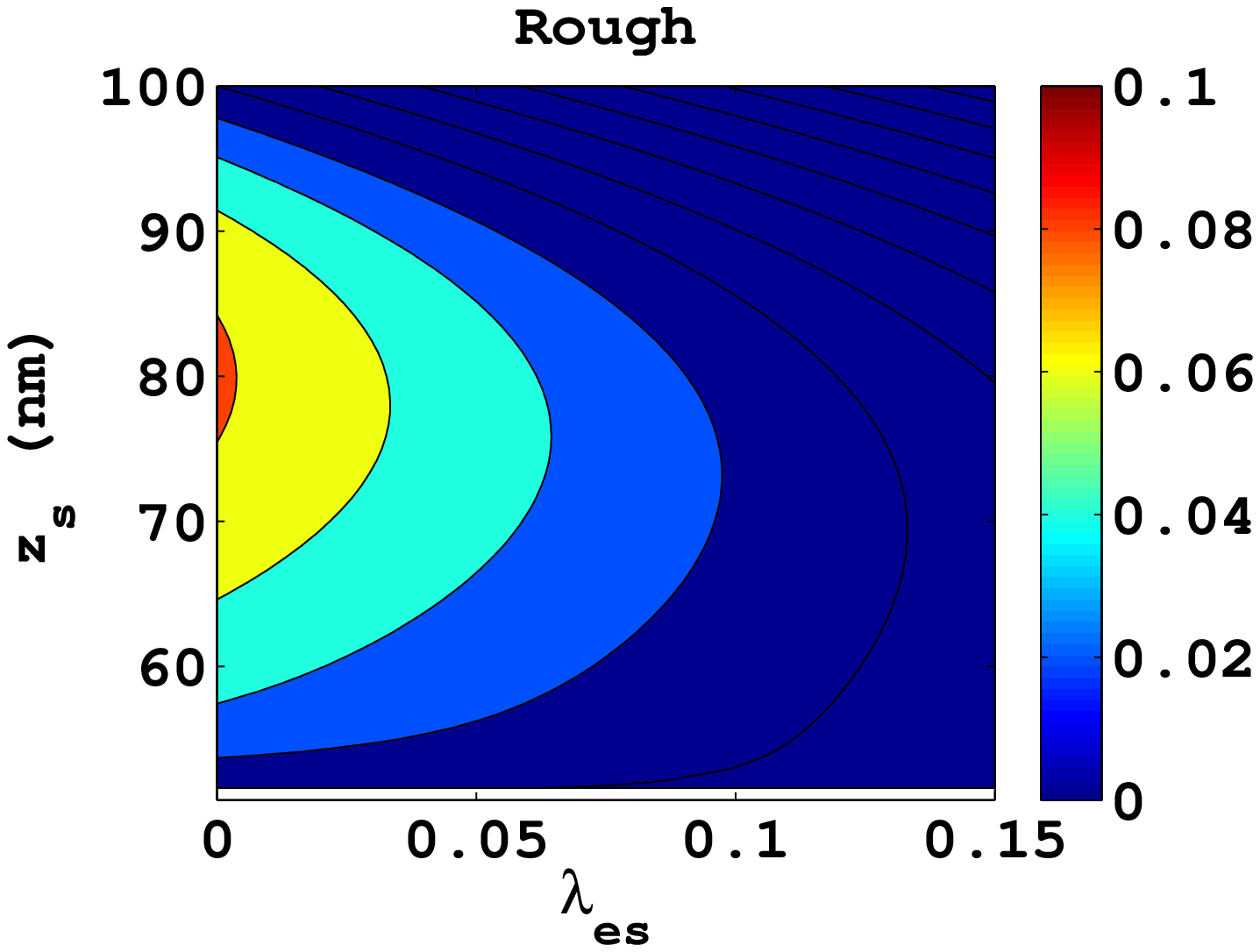}}
\subfloat[][]{\label{fig:fig3d}\includegraphics[width=0.45\textwidth]{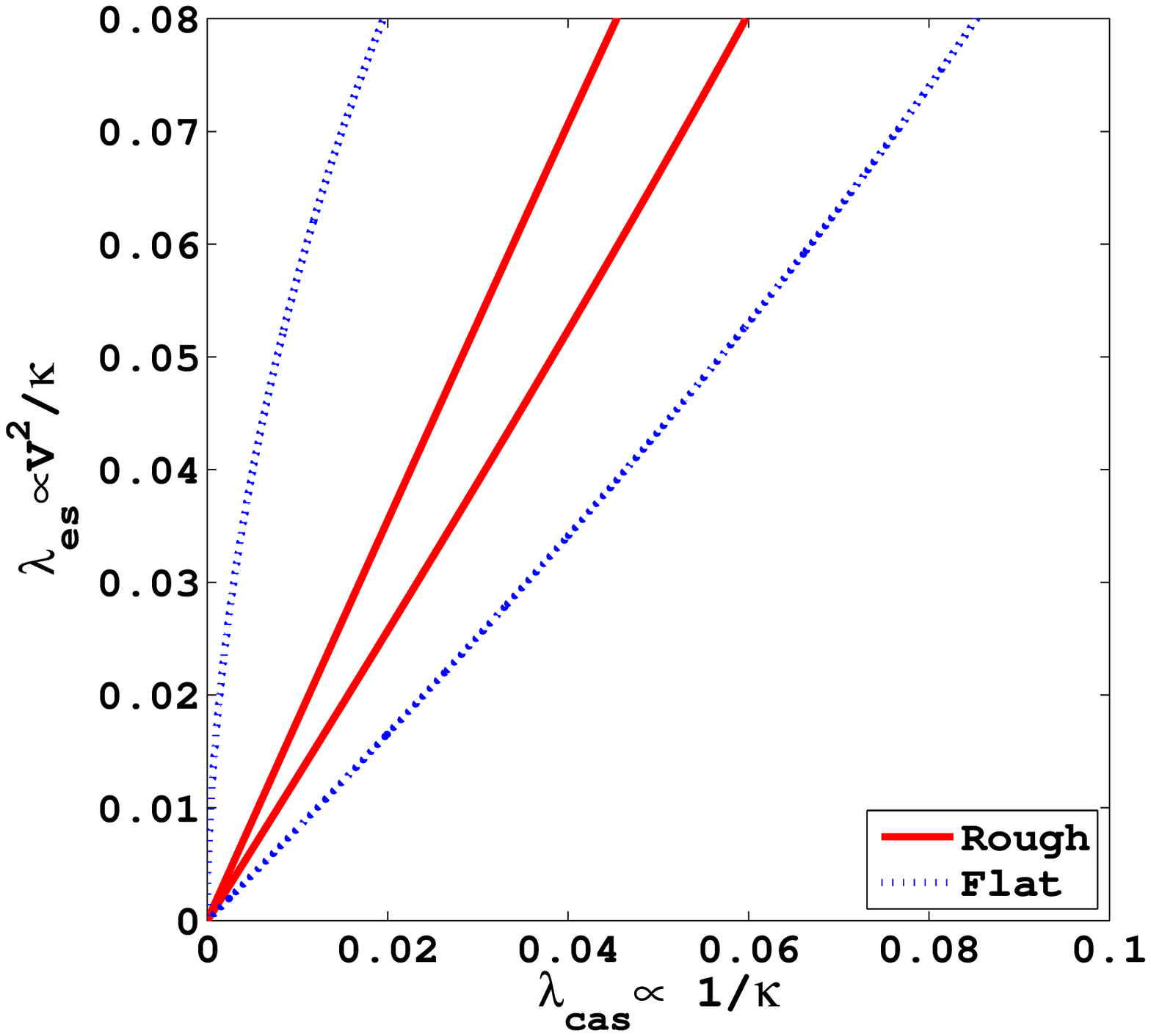}}
	\caption{\protect\subref{fig:fig3a}  3D Bifurcation diagram for a MEMS actuated by both Casimir and electrostatic forces for both a rough and a flat surface. The innermost graph represents the rough case.  \protect\subref{fig:fig3b} Side view of the innermost graph of Fig. \ref{fig:fig3a} \protect\subref{fig:fig3c} Top view of the innermost graph of Fig. \ref{fig:fig3a}. \protect\subref{fig:fig3d} Two parameter bifurcation diagram under the conditions of Eqs. \eqref{eq:3dbif} and \eqref{eq:3dbif2}. In the origin the equilibria join.}	
\label{fig:fig3}
\end{figure*}

Using measured material and surface  properties  and realistic device dimensions,  we  studied the effect of the Casimir-Lifshitz force on the actuation  dynamics of MEMS. It has turned out that this force is equivalent to a voltage between 0.2 and 0.6 V at separations below 100 nm, which is comparable to the electrostatic force typically used to actuate MEMS. We have shown that random roughness has an overall strong effect on the availability of equilibria of MEMS oscillators: the rougher the surface, the lower the maximum values of the bifurcation parameters, and hence the higher the spring constant must be in order for equilibria to exist. Finally, the shift of the minimum separation due to surface roughness moves the stable center curve closer to the unstable saddle point curve in the bifurcation diagram as a function of the stationary points. However, there is a trade-off: a MEMS of which the actuating components have rough surfaces benefits more from a strong spring constant than one with a flat surface.  This is because surface roughness brings the two equilibrium points closer together in terms of the separation, but it prevents the surface forces that give rise to stiction from becoming too large. Most likely the optimum lies somewhere in between a flat surface (with $d_0=0$) and a very rough surface (with $d_0\approx$ 50 nm).  Finally, another effect that deserves further consideration is that of nonlinearity, i.e. deviations from Hooke's law \cite{Aldridge2005,Husain2003} at sufficiently large amplitudes.  For example, for operation of a resonator in vacuum studied here, we could have some manifestation of nonlinear behavior for oscillation amplitudes $\sim$100 nm,\cite{Ekinci2005} which is comparable to the device dimensions, i.e. $L_0$ considered here. Surface roughness could possibly reduce or prevent such nonlinearities, because high surface peaks limit large amplitude oscillations. Our analysis applies to motion in vacuum or dilute gases, where friction losses can be ignored. Qualitatively, the center equilibrium in Fig. \ref{fig:fig2b} becomes an inward `sink' spiral in the presence of friction. A more detailed analysis of the effect of hydrodynamic drag forces\cite{Craig2003, Vinogradova2003, Vinogradova2006} present in air will be combined with that of nonlinearity in a future study.

\begin{acknowledgments}
We thank A. Doelman and H. W. Broer for useful discussions. WB also gratefully acknowledges I. Hoveijn for his help with the bifurcation analysis. 
\end{acknowledgments}

\bibliography{Casimir}

\pagebreak

\pagebreak

\pagebreak

\end{document}